\pdfoutput=1
\documentclass[11pt,a4paper]{article}
\usepackage[utf8]{inputenc}
\usepackage[T1,T5]{fontenc}
\usepackage{lmodern}
\usepackage{microtype}
\usepackage[margin=2.5cm]{geometry}
\usepackage{setspace}
\usepackage{amsmath,amssymb,amsfonts,amsthm}

\usepackage{graphicx}
\usepackage{booktabs}
\usepackage{array}
\usepackage{indentfirst}
\usepackage{placeins}

\usepackage{caption}
\usepackage[hyphens]{url}
\usepackage{xcolor}
\usepackage[authoryear,round]{natbib}
\usepackage[colorlinks=true,urlcolor=blue!60!black,linkcolor=black,citecolor=black]{hyperref}
\title{\textbf{Time-Aware Urban Flood Routing with Sticky Edge Closure on Forecast Depth: A Hanoi Case Study}}

\author{Mai Anh Tran, Vy Nguyen, Vova Nguyen, and Minh Hoa Bui\\[0.5em]
\normalsize TECOTEC Technologies Joint Stock Company, Hanoi, Vietnam\\
\normalsize \texttt{maianh@tecotec.com.vn}}

\date{October 2026}

\begin{document}

\maketitle

\begin{abstract}

During urban flash floods, commuters and guiding agencies require awareness towards which roads remain passable when a vehicle actually arrives, rather than relying solely on which links appear open at initial departure. Most existing flood-routing tools apply a single static snapshot to each route; meaning that corridors that appear clear initially can become flooded before arrival. This paper presents a routing engine that ingests a two-hour water forecast, renewed every 15 minutes, and plans each journey around roads that the forecast closes for that specific vehicle class. Under this formulation, any road that closes mid-trip remains closed for that query, which keeps route calculation fast enough for live requests while operating inside a production GraphHopper routing server. Testing on a recorded storm in western Hanoi revealed that a router checking only departure conditions sent 7.8\% of routes into roads flooded upon arrival. In comparable trips, the proposed rule produced longer routes in 3.95\% of cases, requiring a median of only 9.2 s of additional travel time. Agencies and navigation platforms can employ this engine to schedule trips by vehicle type, allowing operators to establish depth limits and lookahead margins as policy parameters. However, several limitations must be noted, as the water depths originate from a hydrodynamic simulation that was not verified against measured water levels, the rainfall was replayed rather than forecast live, and depth cutoffs reflect policy settings rather than measured physical thresholds.

\end{abstract}


\vspace{1.5em}

\section{Introduction}

\subsection{Flash Floods and Time-Dependent Routing}

Convective monsoon storms in rapidly urbanizing Southeast Asian cities such as Hanoi flood streets within minutes of onset, overwhelming municipal drainage capacity (Luo et al., 2018). In western Hanoi, arterial corridors cross former agricultural lowlands where subsurface drainage lags behind surface road paving, creating sharp disparities in flood susceptibility across adjacent neighborhoods (Nguyen et al., 2026). Given that street water rises rapidly during a commute, water depths along low-lying segments can cross vehicle passability thresholds while a trip is underway. Empirical flood-mobility studies confirm that shallow inundation triggers non-linear travel-time escalations, making trip completion highly sensitive to departure timing (Suwanno et al., 2023). While contemporary hydrodynamic models generate multi-step inundation forecasts across rolling horizons, existing flood routers continue to evaluate road passability against a single frozen snapshot at departure (Alabbad et al., 2024), exposing en-route travelers to floodwater that deepens before arrival.

This reliance on departure-time snapshots characterizes both established methodological routes in flood-transport literature, yet neither approach resolves the temporal mismatch for an active journey. In the first route, accessibility studies and decision support tools map water depth to continuous speed-reduction curves or binary link closures as a static spatial overlay on road networks (Alabbad et al., 2024; Pregnolato et al., 2017). While some studies evaluate time-varying accessibility across sequential storm slices (Miller, 2022) or integrate production engines such as Open Source Routing Machine (Anderson et al., 2025) and GraphHopper (Karanjit et al., 2024), all evaluate passability against a single forecast snapshot per query. In the second route, offline traffic simulators such as MATSim (Bi et al., 2026) and SUMO (Li et al., 2026) inject pre-scripted closures to evaluate aggregate traffic delay rather than routing individual vehicles in real time. Similarly, deep learning models predict multi-hour road inundation states without coupling predictions to trajectory-level path search (Yuan et al., 2022). As Ermagun et al. (2026) emphasize, neither static spatial overlays nor offline traffic microsimulations identify a path whose edges remain passable at the future times when a traveler reaches them.

Evaluating forecast water depth at the projected arrival time of each road segment resolves this en-route arrival mismatch, yet it introduces a fundamental algorithmic barrier when floodwaters recede within the prediction window. In time-dependent shortest-path routing, forward label-setting extensions of Dijkstra's algorithm and A* search maintain polynomial complexity only if every edge traversal function satisfies the First-In, First-Out (FIFO) condition: departing an edge later never yields an earlier arrival at its head vertex (Bast et al., 2016; Kaufman \& Smith, 1993). However, a flooded road segment that closes during peak rainfall and reopens as water drains completely violates this FIFO condition: an early arrival is blocked, whereas a later arrival after recession traverses the segment without interruption. Consequently, shortest-path search becomes NP-hard when waiting is allowed only at designated locations (Kleff et al., 2020). While scheduled freight transport can absorb closures as rest breaks at designated parking facilities (van der Tuin et al., 2018), idling on an inundated street during an active storm is hazardous and unacceptable for urban commuters. Recent prototypes sidestep this difficulty by assuming inundation only worsens during rescue operations (Liang et al., 2025) or by restricting routing to small subgraphs (Jao et al., 2025), leaving the non-FIFO reopening dilemma unsolved on city-scale networks.1.2 Vehicle Limits and Forecast Uncertainty

Beyond the algorithmic challenge of non-FIFO road reopening, translating multi-step depth forecasts into passable corridors requires addressing sharp differences in vehicle vulnerability and upstream prediction error. Laboratory flume experiments and flood hazard guidelines establish a stationary still-water stability threshold of 0.3 m only for small passenger cars (AIDR, 2017; Martínez-Gomariz et al., 2018; Shand et al., 2011; Smith et al., 2014). These guidelines emphasize that commercial trucks, emergency vehicles, and motorcycles lack calibrated hydrodynamic stability curves in published literature (Smith et al., 2014). Nevertheless, prevailing transport studies apply a uniform passenger-car cutoff across mixed traffic (Li et al., 2026) or select regional thresholds across wide ranges under the assumption that aggregate delay is insensitive to the exact cutoff (Kasmalkar et al., 2020). In Southeast Asian cities, two-wheeled motorbikes dominate daily commuting traffic (Chapuis et al., 2022; Ermagun et al., 2026), and traffic volumes drop sharply once street water reaches axle-level depths of 15 to 20 cm (Mamuyac et al., 2024; Nguyen, 2022; Vu, 2026). Mode-specific cutoffs fundamentally change which streets remain open across mixed fleets (Dasallas et al., 2025). Furthermore, drainage capacities vary across Hanoi wards (Nguyen et al., 2026), while official flood bulletins are broadcast retrospectively without predictive horizons or depth metrics (HSDC, 2024; Kefi et al., 2018). Given upstream forecast errors (Murray-Tuite et al., 2017), vehicle depth limits cannot function as physical safety guarantees, serving instead as policy parameters paired with a pre-closure lookahead margin, a shallow-water precaution rule, and an explicit regret sensitivity analysis.

\subsection{This Study}

To resolve the computational non-FIFO barrier while accommodating these policy constraints, this paper introduces a routing engine that forbids a road from reopening inside a single query. Under this sticky closure rule, each road edge receives an effective closing time, defined as the earliest forecast step at or after departure at which predicted depth reaches the vehicle depth limit. The edge remains closed for any projected arrival within a lookahead margin before that step and for every later arrival in the same query. This rule is applied directly to forecasts of a 2D subgrid hydrodynamic model for western Hanoi (Bui \& Tran, 2026) inside a GraphHopper routing engine reading a compact road-edge depth table refreshed on each forecast cycle. By forbidding reopening within a query, the arrival function of every edge remains monotonic, the FIFO condition holds across the network, forward label-setting search stays exact, and no traveler waits on a flooded street. What the rule costs is routing capacity, as a drained road stays closed for the rest of the query, and each road closes slightly before depth reaches the limit. Replaying forecast tables for a convective storm in western Hanoi, this study measures what is sacrificed by comparing the sticky rule against a departure-step snapshot router, a horizon-maximum router, and an exact arrival-only search that permits reopening.

These empirical comparisons are evaluated within specific physical boundaries: they draw on one convective storm replayed across one 31.94 km\textsuperscript{2} urban catchment, where the overland hydrodynamic model contains no subsurface sewers or pumps and vehicle depth limits serve as policy parameters rather than calibrated physical thresholds. Within this defined scope, four principal contributions follow. First, this investigation provides a direct empirical measurement of the cost of forbidding road reopening, decomposed into lookahead margin and closure carry-forward, evaluated against single-timestep, horizon-maximum, and arrival-only benchmarks. Second, it formalizes the sticky closure rule alongside mathematical proofs of the FIFO lemma and the exactness of earliest-arrival and multi-label profile searches, deploying the formulation in a production routing engine reading a compact binary edge-depth table. Third, because calibrated limits for motorbikes and trucks do not exist, the study audits threshold provenance, evaluates the asymmetry of planning above versus below the true limit across a combined cost scale, and frames vehicle limits as policy choices. Fourth, the design establishes an interactive departure-time advisory and a two-option precaution gate that provides a recommended shallow-water detour alongside a flagged through-water route bounded by the vehicle limit.

\section{Theoretical Background and Related Work}

\subsection{Shortest Paths and Road Closures}

Time-dependent route planning extends static shortest-path search by determining the travel time of each road segment based on arrival time rather than assuming conditions remain fixed (Dreyfus, 1969). Dreyfus (1969) observed that evaluating departure-time-dependent arc costs during forward label relaxation generalizes Dijkstra's algorithm without altering its label-setting structure, provided that the network satisfies temporal consistency. Kaufman and Smith (1993) formalized this requirement as the First-In, First-Out (FIFO) consistency condition, where departing an arc later never yields an earlier arrival at its head vertex. Under this condition, a time-dependent fastest path can be computed with the same asymptotic complexity as a static shortest path. Bast et al. (2016) synthesized this non-overtaking property across route planning engines, showing that Dijkstra and A* remain exact on time-dependent road graphs whenever later departures cannot overtake earlier ones. However, when road segments close temporarily and subsequently reopen, the arrival-time function drops from infinity during closure back to a finite arrival time once the closure lifts, violating the non-overtaking condition unless the routing model specifies a waiting policy.

Whether a non-FIFO network can be reduced to an effective FIFO network relies fundamentally on where and how a vehicle may wait. Orda and Rom (1990) analyzed shortest-path routing under unrestricted waiting, forbidden waiting, and source-only waiting, finding that unrestricted waiting at intermediate vertices absorbs unfavorable traversal windows, whereas without intermediate waiting, only unrestricted departure selection at the origin guarantees a simple path with minimal delay. Dean (2004) formalized vertex waiting policies governing allowable durations and travel costs, while Foschini et al. (2011) established that a non-FIFO edge is equivalent to a FIFO edge when unrestricted waiting is permitted at its tail vertex. However, when waiting is prohibited or restricted, this equivalence breaks down: He et al. (2020) demonstrated that penalizing or restricting waiting on non-FIFO networks can render the minimum-arrival-time path problem NP-hard. Kleff et al. (2020) analyzed this barrier for temporary road closures modeled as disjoint intervals, proving that the quickest-path problem becomes NP-hard when waiting is permitted only at designated parking locations. While scheduled freight transport can absorb predictable driving bans as mandatory rest breaks at rest areas (van der Tuin et al., 2018), urban vehicles cannot safely park on an inundated street while floodwaters recede. Consequently, preserving polynomial-time search without roadside waiting requires modifying the closure model so that the FIFO condition holds across all edges, a requirement that a monotone closure rule that forbids mid-query reopening satisfies directly.

A second algorithmic barrier concerns updating speedup structures when environmental forecasts rewrite edge traversal functions every forecast cycle. Batz et al. (2013) introduced Time-Dependent Contraction Hierarchies, and Strasser et al. (2021) developed customizable time-dependent shortcut hierarchies that store paths rather than travel-time functions. Both hierarchy families require costly customization passes whenever underlying arc functions change. In dynamic environments, Delling and Nannicini (2012) proposed core routing methods updating piecewise-linear costs with modest effort, while Baum et al. (2016) developed a dynamic routing algorithm integrating global metric updates and user restrictions. Furthermore, Werner and Zeitz (2022) demonstrated that A* guided by lower-bound potentials incorporates live traffic updates within fractions of a minute without rebuilding contraction hierarchies. For urban networks where hydrodynamic forecasts update thousands of road edges each cycle and closure thresholds vary by vehicle class, an uncontracted forward search over flat closure tables avoids shortcut customization overhead, provided the closure rule preserves FIFO consistency.

\subsection{Flood Routing and Vehicle Limits}

Existing flood-transport studies translate surface water depth into road network impedance primarily through continuous speed-reduction curves or binary link closures on static flood maps. Pregnolato et al. (2017) synthesized video observations and experimental data into a depth-disruption curve relating water depth to vehicle speed up to a 30 cm passenger-car ceiling, which Liu et al. (2022) applied in commuting simulations for Wuhan to block passenger cars once depth exceeded the 30 cm engine air-intake level. In emergency accessibility and regional risk modeling, Coles et al. (2017) overlaid 25 cm depth masks onto York's road network, Miller (2022) evaluated accessibility across sequential storm slices as independent static networks, and Alabbad et al. (2024) noted that web tools for flood routing remain scarce and continue to rely on static scenario overlays. Similarly, studies integrating production open-source engines follow this static pattern: Anderson et al. (2025) removed links reaching 300 mm in Open Source Routing Machine, and Karanjit et al. (2024) coupled flood predictions to GraphHopper using static avoidance polygons for a single forecast snapshot, without evaluating water depth at projected edge arrival times.

To capture temporal flood evolution more realistically, a second stream of research couples hydrodynamic models to offline traffic microsimulators or custom emergency dispatch prototypes. Pyatkova et al. (2019) coupled 2D flood simulations with SUMO, demonstrating that delays persist after floodwaters subside, while Dave et al. (2026) combined a 0.3 m closure cutoff with shallow-water speed reductions. Bi et al. (2026) injected time-varying speed adjustments and closures into MATSim through pre-scripted network change events, and Li et al. (2026) updated closures dynamically inside SUMO using its built-in Dijkstra router. For emergency evacuation, He et al. (2021) developed a cellular-automata route optimization model incorporating hydrodynamics and human response delays, Xiao et al. (2025) demonstrated that periodic replanning under a 30 cm closure rule achieves substantial travel-time reductions, Liang et al. (2025) evaluated arc speeds at tail-node arrival times under worsening inundation, and Jao et al. (2025) formulated flood evacuation on a 4,892-edge network as a time-dependent arc routing problem over one-minute intervals. However, none of these frameworks resolves the non-FIFO violation that arises when floodwater recedes within a query horizon, nor measures what a monotone closure rule sacrifices relative to an arrival-only dynamic router that permits mid-query reopening.

Translating predicted water depths into closures exposes an evidentiary gap across vehicle classes. Laboratory flume campaigns and flood hazard guidelines provide stationary stability criteria primarily for passenger cars and pedestrians. Shand et al. (2011), Smith et al. (2014), and Martínez-Gomariz et al. (2018) established a 0.3 m limiting still-water depth for small passenger cars, codified by AIDR (2017) as hazard category H1. However, Smith et al. (2014) emphasized that commercial trucks, emergency vehicles, and motorcycles represent unmeasured gaps in physical flume testing. Furthermore, vehicle and pedestrian stability depend jointly on water depth and flow velocity (Arrighi et al., 2017; Defra \& Environment Agency, 2006; Martínez-Gomariz et al., 2017), with heavy-vehicle wading heights substantially exceeding those of passenger cars (Contreras-Jara et al., 2018). Separating pedestrian and vehicle thresholds also noticeably alters which corridors are obstructed at flood peak (Dasallas et al., 2025). Transport simulations nonetheless frequently collapse closures to a single uniform cutoff across all traffic (Kasmalkar et al., 2020; Li et al., 2026). This simplification is especially restrictive in Southeast Asian cities where motorbikes dominate daily commuting (Chapuis et al., 2022; Ermagun et al., 2026). Motorbikes experience severe speed reductions in shallow water (Nguyen, 2022), traffic volumes drop sharply at 20 cm depth (Mamuyac et al., 2024), and municipal traffic police advise riders to halt once water reaches 20 to 30 cm (Vu, 2026). Given that calibrated hydrodynamic curves do not exist for motorbikes, depth limits applied to them must be treated as policy parameters rather than physical stability thresholds.

When rising floodwaters block corridors or degrade network speeds, adjusting the departure time at the origin provides an alternative to spatial detouring. Orda and Rom (1990) showed that unrestricted departure selection at the source vertex can eliminate intermediate waiting, while Batz et al. (2013) demonstrated how profile queries characterize arrival time as a function of departure time. In empirical flood-mobility research, Suwanno et al. (2023) demonstrated that trip durations escalated non-linearly from 40 to 670 minutes as flood depths rose from 5 to 30 cm, highlighting extreme sensitivity to departure timing. Furthermore, Ermagun et al. (2026) identified departure-time adjustment and trip cancellation as primary behavioral responses to urban flooding that remain poorly integrated into existing flood routers. Existing dynamic evacuation frameworks model departure timing from a centralized scheduling or scenario-testing perspective rather than as an interactive per-trip query (Bahrami et al., 2026; Bi et al., 2026; He et al., 2021; Pel et al., 2008). An individual commuter whose usual route is blocked at departure requires an interactive routing service that scans subsequent forecast steps within the active prediction window and reports the earliest step at which a passable path opens.

\subsection{Forecasts and Routing Uncertainty}

The design of an online time-dependent flood router is bounded by the temporal horizon and predictive accuracy of the upstream inundation forecast. Pluvial flood disruption assessments typically operate across characteristic timescales of 60 to 200 minutes (Rebally et al., 2021), matching the multi-step lookahead window of an urban trip. While spatio-temporal deep learning models (Yuan et al., 2022) and coupled hydrodynamic digital twins refresh depth grids on rolling cycles of a few minutes, consuming these forecasts in real-time routing engines (Baum et al., 2016; Werner \& Zeitz, 2022) requires an explicit posture toward forecast error. Pre-trip routes planned on deterministic forecasts remain vulnerable to timing and depth errors in the predicted flood surface (Gao \& Chabini, 2002). In an empirical evaluation, Murray-Tuite et al. (2017) observed an 8.92\% false-positive rate and a 31.50\% false-negative rate in flood hazard predictions used for traffic clearing and rerouting. In urban hydrodynamic simulations constructed without complete subsurface pipe data, false negatives and premature recession estimates can expose travelers to flooded streets if the router trusts mid-trip road reopening or treats vehicle depth cutoffs as exact physical thresholds.

These literature findings reveal a joint gap across theory, modeling, and empirical validation. Time-dependent closure theory shows why reopening roads makes shortest-path search hard and why a monotone closure rule restores polynomial tractability, whereas flood-routing studies apply static snapshots or offline simulations without measuring what a monotone rule costs against an arrival-only router that permits reopening. Furthermore, vehicle-limit evidence is absent for the dominant local travel mode, and forecast uncertainty is rarely propagated into route validity. These gaps motivate three design elements: a road-edge depth table refreshed on each forecast cycle paired with a monotone sticky closure rule and a lookahead margin; an interactive departure-time advisory and two-option precaution gate; and an asymmetric regret evaluation across vehicle classes. Measuring the price of the monotone sticky closure rule against dynamic benchmarks is the question that the empirical evaluation answers.

\section{Description of the Case Study and Available Data}

\subsection{Study Area and Event}

The case study examines a rapidly urbanizing catchment in western Hanoi, Vietnam (Bui \& Tran, 2026; Nguyen et al., 2026), spanning the suburban-to-urban transition across the My Dinh and Tay Mo corridors, represented by the 2D hydrodynamic domain covers an area of 31.94 km\textsuperscript{2}. Across this sector, high-capacity radial expressways intersect former agricultural lowlands and dense village settlements where subsurface storm drainage has lagged behind surface road paving, creating sharp disparities in flood susceptibility across adjacent neighborhoods (Nguyen et al., 2026). Extreme convective storms rapidly overwhelm local drainage and produce street ponding up to 0.5 m within minutes (Luo et al., 2018). However, municipal flood reporting in Hanoi remains retrospective (HSDC, 2024; Kefi et al., 2018), broadcasting alerts only after key intersections are flooded, leaving commuters without predictive horizons or street-level depth severity estimates. To model travel across this network, the street network was extracted from OpenStreetMap into the GraphHopper routing engine, comprising 67,833 road edges and 51,328 vertices. Each bidirectional or one-way street segment between intersection or splitter vertices forms one edge. To ensure index stability across repeated graph builds, each edge receives a composite identifier pairing its OpenStreetMap way identifier with a zero-based segment index.

Meteorological forcing is derived from the September 16 to 18, 2026 convective storm over Hanoi, captured by the NASA IMERG Late Run half-hourly 0.1° precipitation product (Huffman et al., 2023). The wettest 12-hour core accumulates 53.88 mm from 23:00 on September 16 to 11:00 on September 17, resampled onto 48 quarter-hour pipeline slots. In the experimental replays, each execution cycle reads the upcoming two hours of this recorded satellite series forward in time, providing a fixed two-hour forecast window without dynamic forecast revisions. Surface runoff and street inundation are simulated using a 2D subgrid hydrodynamic model in the 3Di framework (Bui \& Tran, 2026). The model operates as an overland representation without subsurface sewer pipes or pumping stations, where simple infiltration (domain mean 11.8 mm/day, median 7.75 mm/day) acts as the sole sink. Given that this infiltration rate removes less than 1 mm over two hours, simulated depth reductions reflect lateral redistribution rather than pipe drainage, causing water to recede much more slowly than in a fully drained urban catchment. To evaluate time-dependent routing, the simulation produces eight forecast depth steps at 15-minute intervals spanning 15 to 120 minutes ahead, matching the characteristic timescale of urban pluvial flood disruptions (Rebally et al., 2021). The evaluation draws on two replay datasets: 45 fragmented binary road-edge depth tables separated by a four-slot pipeline gap, and 57 continuous consecutive quarter-hour runs. Eleven runs from the continuous sequence, stratified across four representative storm stages, serve as the primary testbed for routing evaluations.

\subsection{Vehicle Profiles and Depth Limits}

Translating road-edge water depths into mode-specific passability rules requires clear physical thresholds. The routing configuration assigns a hard closure depth limit $L$ and a shallow-water precaution threshold $L_{\text{prec}} = f \cdot L$ (with default fraction $f = 0.50$) across four distinct transport profiles. Specifically, motorbikes receive $L = 20\text{ cm}$ ($L_{\text{prec}} = 10\text{ cm}$), passenger cars receive $L = 30\text{ cm}$ ($L_{\text{prec}} = 15\text{ cm}$), commercial trucks receive $L = 35\text{ cm}$ ($L_{\text{prec}} = 17.5\text{ cm}$), and pedestrians receive $L = 40\text{ cm}$ ($L_{\text{prec}} = 20\text{ cm}$). Across published hydrodynamic stability studies and flood hazard manuals, only the 30 cm passenger-car limit corresponds directly to a published stationary stability criterion (Shand et al., 2011; Smith et al., 2014; AIDR, 2017; Martínez-Gomariz et al., 2018). In these guidelines, small cars are bounded by $D \cdot V \le 0.3\text{ m}^2/\text{s}$ and a 0.3 m limiting still-water depth, aligning with the 30 cm speed-disruption ceiling in Pregnolato et al. (2017). For motorbikes, the dominant personal transport mode in Hanoi (Chapuis et al., 2022), manufacturer manuals specify ground clearances of 141 to 142 mm alongside 25 cm axle heights. Furthermore, traffic surveys document sharp reductions in motorcycle volumes at 20 cm depth (Mamuyac et al., 2024; Nguyen, 2022), and municipal traffic police advise riders to halt when water reaches 20 to 30 cm (Vu, 2026). Setting the motorbike closure limit at 20 cm blocks edges at the lower boundary of this official halt advisory, with the 10 cm precaution threshold penalizing shallow water before the 15 cm axle level. For commercial trucks, the 35 cm threshold represents an engineering assumption sitting above small cars but below heavy-vehicle wading heights of 110 to 150 cm (Contreras-Jara et al., 2018; Defra \& Environment Agency, 2006). For pedestrians, the 40 cm limit is a conservative assumption set below the 0.5 m child and 1.2 m adult low-hazard thresholds (Cox et al., 2010; Smith et al., 2014). All four thresholds are treated strictly as routing policy parameters rather than absolute physical safety guarantees.

\begin{figure}[htbp]
\centering
\includegraphics[width=0.9\textwidth]{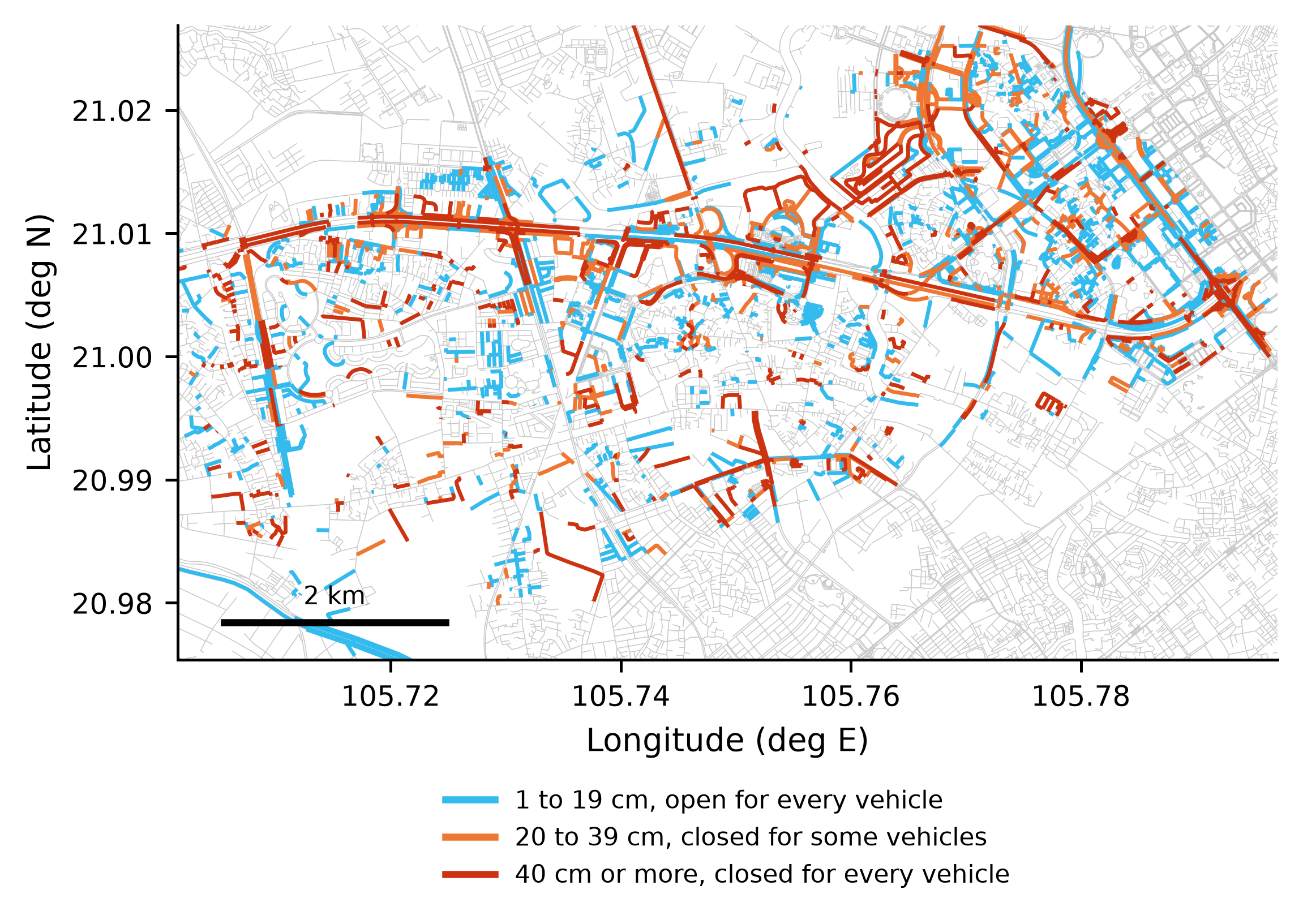}
\caption{Figure 3.1: Road edges of the study area that are wet at the step with the most wet edges of the wettest run of the continuous replay, colored by water depth. Edges that are dry are drawn in grey. The bands of 1 to 19 cm, 20 to 39 cm, and 40 cm or more follow the vehicle limits of 20 cm to 40 cm, so an edge in the middle band is closed for some vehicles and an edge in the last band is closed for every vehicle.}
\end{figure}

\section{Methods}

\subsection{Flood Depth Sampling}

The routing engine consumes one object per forecast run: a table recording the water depth of every road edge at every forecast step. This table is constructed from the depth rasters of the hydrodynamic forecast (Bui \& Tran, 2026) in two stages: centerline sampling and binary serialization. During sampling, each road edge in the routing graph is segmented at 1 m intervals along its centerline. The midpoint of each interval is matched to the hydrodynamic raster cell containing it. The depth of the edge at that forecast step is assigned as the maximum depth observed among all its 1 m samples, rounded to whole centimetres, while an edge touching no wet cell reads zero. Sampling occurs once per forecast step, and a single depth value serves both travel directions because each bidirectional street segment is represented as one edge.

The maximum depth is selected over an average because a vehicle traversing an edge encounters its deepest inundated point, whereas an average dilutes localized ponding along extended links. To quantify how these statistics diverge, three metrics were computed for every edge with at least one wet sample at the peak-inundation step of the wettest continuous replay run: the maximum, the mean over all samples (counting dry samples as zero), and the mean over wet samples only. The maximum depth closes 1,922 edges under the motorbike limit of 20 cm, whereas the all-sample mean reopens 1,044 of them (54.3\%), with reopened shares rising to between 62\% and 64\% for car, truck, and pedestrian limits (Table 4.1). Even the wet-sample mean reopens between 351 and 636 edges per vehicle class.

\begin{table}[htbp]
\centering
\caption{Table 4.1: Edges closed by the per-edge maximum depth and reopened by a mean, at the step with the most wet edges in the wettest run of the continuous replay. Percentages use the number of edges closed by the maximum.}
\resizebox{\textwidth}{!}{%
\begin{tabular}{l l l l l l}
\toprule
Vehicle & Limit (cm) & Closed by maximum & Closed by all-sample mean & Reopened by all-sample mean & Reopened by wet-sample mean \\
\midrule
Motorbike & 20 & 1,922 & 878 & 1,044 (54.3\%) & 636 (33.1\%) \\
Car & 30 & 1,303 & 494 & 809 (62.1\%) & 567 (43.5\%) \\
Truck & 35 & 1,052 & 390 & 662 (62.9\%) & 449 (42.7\%) \\
Foot & 40 & 868 & 314 & 554 (63.8\%) & 351 (40.4\%) \\
\bottomrule
\end{tabular}%
}
\end{table}

Sampled depths are serialized into a compact binary file per forecast run, allowing both the routing engine and departure advisory service to read identical bytes via memory mapping without database overhead. The file opens with a fixed header recording format markers, version numbers, step counts, edge counts, the SHA-256 hash of the routing graph, and run identifiers. Step times follow as 64-bit Unix timestamps. The payload stores one unsigned 16-bit integer depth in centimetres per edge and step, ordered by edge index so that row position defines the edge identifier. The engine verifies graph hashes upon load to prevent version skew between network topology and depth data. The binary table carries no vehicle limits; consuming services apply mode-specific policy limits dynamically.

\subsection{Sticky Closure}

Depth forecasts are available at discrete step times $s_0 < s_1 < \dots < s_{K-1}$. For a trip requested at time $t_{\text{req}}$, the active forecast window begins at $t_w$, defined as the latest step time not exceeding $t_{\text{req}}$. Let $d_e(k)$ be the depth of edge $e$ at step $k$, let $L$ be the vehicle depth limit, and let $\tau_e \ge 0$ denote edge travel time. Under the sticky closure rule, a traveler entering edge $e$ at time $a$ finds it closed if any step $k$ with $t_w \le s_k \le a + m$ satisfies $d_e(k) \ge L$. Here, $m = 600\text{ s}$ serves as a lookahead margin that closes edges to travelers entering within 10 minutes before forecast depth reaches the limit. An edge closure never lifts within the duration of a query, establishing the monotone property of sticky closure.

Equivalently, define the effective closing time of edge $e$ as $c(e) = \min \{ s_k \mid s_k \ge t_w \text{ and } d_e(k) \ge L \}$, with $c(e) = +\infty$ if no qualifying step exists. An edge is open at entry time $a$ if and only if $a < c(e) - m$. The arrival function specifying departure from the head vertex is formulated as:

\begin{equation}
f_e(a) = \begin{cases} a + \tau_e & \text{if } a < c(e) - m \\ +\infty & \text{otherwise} \end{cases}
\end{equation}

Given that $c(e)$ depends solely on the forecast window and vehicle limit, it is invariant to arrival time during a query.

Lemma 1 (FIFO condition): For every edge $e$ and every pair of entry times $a \le b$, the arrival function satisfies $f_e(a) \le f_e(b)$ and $f_e(a) \ge a$.

To prove this condition, consider two cases determined by the effective closing time $c(e)$ and lookahead margin $m$. If $b < c(e) - m$, then entry time $a < c(e) - m$ as well, yielding $f_e(a) = a + \tau_e \le b + \tau_e = f_e(b)$. If instead $b \ge c(e) - m$, then $f_e(b) = +\infty \ge f_e(a)$. The lower bound condition $f_e(a) \ge a$ follows directly from non-negative edge traversal times $\tau_e \ge 0$, establishing that entering an edge later never yields an earlier arrival at its head vertex.

Composing these edge functions along a path establishes two key structural properties. First, path arrival times are non-decreasing with departure time, proving that intermediate vertex waiting cannot improve arrival times (Dean, 2004; Foschini et al., 2011). Second, any path containing cycles can drop intermediate loops without delaying arrival, ensuring that optimal earliest-arrival paths are simple paths.

Theorem 1 (Earliest-arrival search): A forward Dijkstra search ordering labels by arrival time and relaxing edge $e$ when open at arrival time $a < c(e) - m$ returns no route if no flood-free path exists, and otherwise returns a path minimizing arrival time among feasible routes.

This earliest-arrival optimality is proved by induction on settled vertex keys. Let $D(v)$ be the minimum arrival time at vertex $v$ across feasible paths entering no closed edge, where priority queue keys represent end times of valid routes and never fall below $D(v)$. Suppose for contradiction that vertex $v$ is settled with a key strictly exceeding $D(v)$. Along an optimal path reaching $v$ at $D(v)$ with non-decreasing entry times, let $u$ be the last settled vertex whose settled time does not exceed its entry time. When $u$ was settled, the outgoing edge along the optimal path was relaxed at a time not exceeding its entry time; by Lemma 1, this edge was open and pushed a heap entry for the successor vertex with key at most $D(v)$. Popping $v$ with a key strictly exceeding $D(v)$ violates the minimum heap invariant, proving that every settled vertex carries its exact least arrival time.

With a Fibonacci heap, the search executes in $\mathcal{O}(|E| \cdot K + |V| \log |V|)$ time, matching static Dijkstra complexity plus $K$ step lookups per edge.

Theorem 2 (Multi-label profile search): When edges and turns carry non-negative profile weights and the search runs to completion without hitting expansion caps, the first label popped at the destination minimizes profile weight among all feasible routes.

The correctness of this multi-label search follows from the dominance criteria applied to candidate paths. Candidate routes are ranked by profile weights where each label holds a pair of weight and elapsed time. A label is dominated and pruned only if another label at the same traversal (same incoming edge, ensuring identical turn costs) has no larger weight and no later elapsed time. By Lemma 1, any continuation feasible from the dominated label is feasible from the dominating label with identical non-negative turn and edge weights, which guarantees that the search preserves optimality.

\begin{table}[htbp]
\centering
\caption{Table 4.2: Step-boundary cases of the sticky closure rule.}
\resizebox{\textwidth}{!}{%
\begin{tabular}{l l l}
\toprule
Case & Behavior of the rule & Reference and engine \\
\midrule
Arrival exactly on a step & The step counts, because $s_k = a \le a + m$ & Identical \\
Window end exactly on a step & The end is included, so a step at $a + m$ closes the road & Identical \\
Departure between two steps & Steps before the window start are ignored and every later step inside the window counts & The reference takes the window start as given. The engine sets it to the step that contains the requested departure, and the reference is called with that value for comparison \\
Depth equal to the limit & The road closes, because the test is $d_e(k) \ge L$ & Identical \\
Water drained before departure & The water is ignored, because steps before the window start are skipped & Identical \\
\bottomrule
\end{tabular}%
}
\end{table}

\subsection{Routing Baselines and Departure Search}

Two routing architectures consume depth data through distinct mechanisms. The single-timestep fallback represents conventional flood-routing practice: it reads edge depths at the departure step alone, computes polygon buffer overlays (buffering edges by 22 m with priority multipliers of 0.2 for 50-75\% limit, 0.05 for 75-100\%, and 0.0 for $\ge 100\%$), and injects spatial avoidances into GraphHopper. This spatial geometry approximates closures without time-awareness. In contrast, the time-aware engine evaluates the binary table under the sticky closure rule via memory-mapped lookups. It executes a two-option precaution search: it plans first under the hard limit $L$; if the path encounters water exceeding the precaution threshold $L_{\text{prec}} = f \cdot L$ ($f = 0.50$ default), a second search runs under $L_{\text{prec}}$. If feasible, the precaution path is returned as the recommended route with the hard-limit path as secondary; otherwise, the hard-limit path is returned flagged as traversing shallow water.

The departure advisory serves travelers whose usual commute route is inundated at departure. The service extracts the baseline route from GraphHopper without flood models, evaluating each constituent edge across upcoming forecast steps with link entry times offset by cumulative travel times. A candidate departure is classified as clear if all edges remain open, blocked if any edge closes on entry, and unknown if the trip extends beyond the forecast horizon, identifying the earliest departure opening the corridor.

\subsection{Experimental Setup}

The experimental protocol couples two testing environments: the deployed GraphHopper stack (incorporating one-way constraints, road classes, and turn costs) and an offline Python reference that evaluates road topology at a uniform 30 km/h free-flow speed. Origin-destination pairs are sampled within the 31.94 km\textsuperscript{2} catchment across three distance strata: short (<2 km), medium (2-5 km), and long (>5 km). Planned routes are evaluated by an independent trajectory checker that re-walks each link at projected arrival times under three step-reading rules: step-at-or-before (latest step not after arrival), nearest-step (closest step in time), and worse-of-two (deeper of the two bracketing steps). The 600 s margin covers the first two rules by construction, while the worse-of-two rule provides an independent stress test.

Conservatism of the sticky rule is evaluated against an exact arrival-only reference search that permits mid-query reopening without edge memory. Since mid-query reopening violates FIFO, the arrival-only search executes an A* label search over simple paths using a 3,000,000 expansion budget, with admissible heuristics derived from vertex waiting relaxations. Routes are classified as identical, longer under sticky closure, or blocked under sticky closure where the reference finds a path. Attribution analysis separates conservatism into lookahead margin effects ($m = 600\text{ s}$ versus $m = 0\text{ s}$) and closure carry-forward. Asymmetric regret is evaluated across a 4x4 matrix crossing planned limits against judging limits, complemented by an exploratory composite cost function combining violation rates, unroutable shares, and detours across 25 parameter weights. Confidence intervals are computed via cluster bootstrapping (5,000 resamples of origin-destination pairs within distance strata).

Evaluations are conducted across two storm sequences: a fragmented sequence of 45 tables separated by a four-slot pipeline gap, and a continuous sequence of 57 consecutive quarter-hour runs spanning cold onset, 1-hour rising, 3-hour mature ponding, and late recession stages. Four routing models are benchmarked: single-timestep departure snapshot, horizon-maximum static closure, naive time-dependent search, and sticky closure. Lookahead sensitivity is analyzed via a margin sweep crossing margins of 0, 300, 450, 600, and 900 s with departure offsets within steps. Reopening prevalence is tracked via edge-level census across all 57 tables, separated from baseline standing water via a permanent-water mask. An adversarial stress simulation with enhanced soil infiltration generates rapid drainage across 10 hours, testing sticky conservatism in an environment where roads reopen far more frequently than in baseline tables. Finally, server latency is recorded across 30-request batches for stock GraphHopper, polygon penalty models, cached backend lookups, and the time-aware binary engine under matched server and local wet-step container loads.

\section{Results}

\subsection{Single-Timestep Violations}

Evaluating planned routes against evolving floodwater shows that conventional static flood routing frequently directs vehicles into inundated streets. Across 3,911 journeys planned by the polygon penalty model under the step-at-or-before rule, 7.8\% of routes (307 trips; 95\% bootstrap interval 4.06\% to 12.67\%) enter road links that are closed at or above the vehicle limit upon arrival. Violation rates peak during the mature ponding stage at 13.0\%, with two-wheelers (10.4\%) and passenger cars (10.8\%) experiencing more than double the failure rates of commercial trucks (4.9\%) and pedestrians (5.9\%). Furthermore, failure rates scale with travel distance, rising from 5.4\% for trips under 2 km to 10.3\% for trips exceeding 5 km, while remaining consistent across departure steps.

\begin{table}[htbp]
\centering
\caption{Table 5.1: Single-timestep routes that enter a closed road on arrival, under the step-at-or-before rule, planned by the polygon penalty model on the deployed GraphHopper stack. The interval is a pair-level 95\% bootstrap interval from 20 pairs per length class, and each group uses its planned routes as the denominator.}
\resizebox{\textwidth}{!}{%
\begin{tabular}{l l l l}
\toprule
Group & Planned routes & Violating routes & Rate \\
\midrule
All routes & 3,911 & 307 & 7.8\% (95\% interval 4.06\% to 12.67\%) \\
Cold onset & 423 & 30 & 7.1\% \\
1.5 h after rain onset & 402 & 15 & 3.7\% \\
3.5 h after rain onset & 376 & 49 & 13.0\% \\
Late recession & 2,710 & 213 & 7.9\% \\
Motorbike & 862 & 90 & 10.4\% \\
Car & 947 & 102 & 10.8\% \\
Truck & 897 & 44 & 4.9\% \\
Foot & 1,205 & 71 & 5.9\% \\
Under 2 km & 1,337 & 72 & 5.4\% \\
2 km to 5 km & 1,170 & 91 & 7.8\% \\
Over 5 km & 1,404 & 144 & 10.3\% \\
Departure step 0 & 1,956 & 155 & 7.9\% \\
Departure step 1 & 1,955 & 152 & 7.8\% \\
\bottomrule
\end{tabular}%
}
\end{table}

Timing and depth distributions indicate that single-timestep failures occur rapidly and under shallow excess water. Across all 307 violating journeys, the median elapsed travel time to the first closed edge is 231 s, with 90\% of violations occurring within 642 s of departure. The median depth exceedance beyond vehicle passability limits is 5.0 cm (mean 7.4 cm, 90th percentile 16.0 cm, maximum 38.0 cm). For motorbikes and passenger cars, median exceedances are 5.0 cm and 4.0 cm, respectively.

\subsection{Baseline Comparison}

In contrast to static routing, walking planned trajectories of the deployed sticky engine over depth tables yields zero entries into closed road segments under both step-at-or-before and nearest-step rules across all 12,785 evaluated routes (Table 5.2). Under the worse-of-two rule, which is not covered by the 600 s margin, the engine records 69 violations out of 3,760 routes (1.8\%) during early storm stages and 2 violations out of 9,025 routes (0.02\%) in late recession.

\begin{table}[htbp]
\centering
\caption{Table 5.2: Verification of the deployed sticky engine by walking each planned route over the depth tables. The margin covers the first two reading rules and not the third. The three early stages and the late stage come from separate runs of the deployed engine, the second in a local container.}
\resizebox{\textwidth}{!}{%
\begin{tabular}{l l l l l l}
\toprule
Storm stage & Planned routes & Entries into a closed road under the sticky rule & Violations, step at or before & Violations, nearest step & Violations, worse of two \\
\midrule
Cold onset & 1,292 & 0 & 0 & 0 & 25 \\
1.5 h after rain onset & 1,253 & 0 & 0 & 0 & 33 \\
3.5 h after rain onset & 1,215 & 0 & 0 & 0 & 11 \\
Three early stages & 3,760 & 0 & 0 & 0 & 69 \\
Late recession, eight tables & 9,025 & 0 & 0 & 0 & 2 \\
\bottomrule
\end{tabular}%
}
\end{table}

Benchmarking four routing models on identical requests over the offline reference graph confirms these safety and detour differences. The single-timestep planner produces a 7.8\% violation rate under the worse-of-two rule (stage-balanced rate 8.5\%), and naive time-dependent search produces 4.2\% violations. The horizon-maximum planner avoids violations completely (0.0\%) but increases mean travel times by 18.4\% and leaves 2.1\% of requests unroutable. Sticky closure records 0.0\% violations under step-at-or-before and nearest-step rules, a 0.7\% violation rate under the worse-of-two rule, a 4.1\% mean travel-time increase, and a 0.12\% unroutable rate.

\subsection{Conservatism of Sticky Closure}

Evaluating sticky closure against the exact simple-path arrival-only reference search on the offline reference graph demonstrates high route agreement. Across 2,318 settled queries, sticky closure returns the exact arrival-only path in 95.9\% of requests, yielding longer routes in 3.95\% of queries (95\% bootstrap interval 2.66\% to 5.28\%, median extra travel time 9.2 s) and blocked routes in 0.12\%. Route deviations concentrate in medium and long distance classes, whereas trips under 2 km exhibit 0.00\% longer routes and 0.00\% blocked routes.

\begin{table}[htbp]
\centering
\caption{Table 5.3: Sticky closure against the simple-path arrival-only reference on the offline reference graph, over exact searches. Shares use the searches in which the arrival-only search finds a route. Intervals are pair-level 95\% bootstrap intervals from 10 pairs per length class (30 pairs).}
\resizebox{\textwidth}{!}{%
\begin{tabular}{l l l l l}
\toprule
Group & Exact searches (share of all) & Sticky longer & Sticky no route & Median extra time (s) \\
\midrule
All & 99.35\% & 3.95\% (2.66\% to 5.28\%) & 0.12\% (0.00\% to 0.34\%) & 9.2 (6.6 to 16.6) \\
Motorbike & 98.81\% & 5.07\% (3.18\% to 7.10\%) & 0.25\% & 11.9 \\
Car & 99.11\% & 3.53\% (2.31\% to 4.82\%) & 0.08\% & 8.3 \\
Truck & 99.70\% & 4.07\% (2.45\% to 5.87\%) & 0.08\% & 10.3 \\
Foot & 99.78\% & 3.22\% (2.13\% to 4.36\%) & 0.08\% & 9.2 \\
Under 2 km & 100.00\% & 0.00\% & 0.00\% & not defined \\
2 km to 5 km & 99.67\% & 3.98\% (1.53\% to 6.87\%) & 0.25\% & 16.6 \\
Over 5 km & 98.39\% & 7.97\% (4.91\% to 10.90\%) & 0.12\% & 8.9 \\
\bottomrule
\end{tabular}%
}
\end{table}

The mechanism producing longer routes is illustrated in Figure 5.1, where the sticky rule detects rising water within the lookahead window and preemptively detours around an edge that the arrival-only search enters shortly before closure.

\begin{figure}[htbp]
\centering
\includegraphics[width=0.9\textwidth]{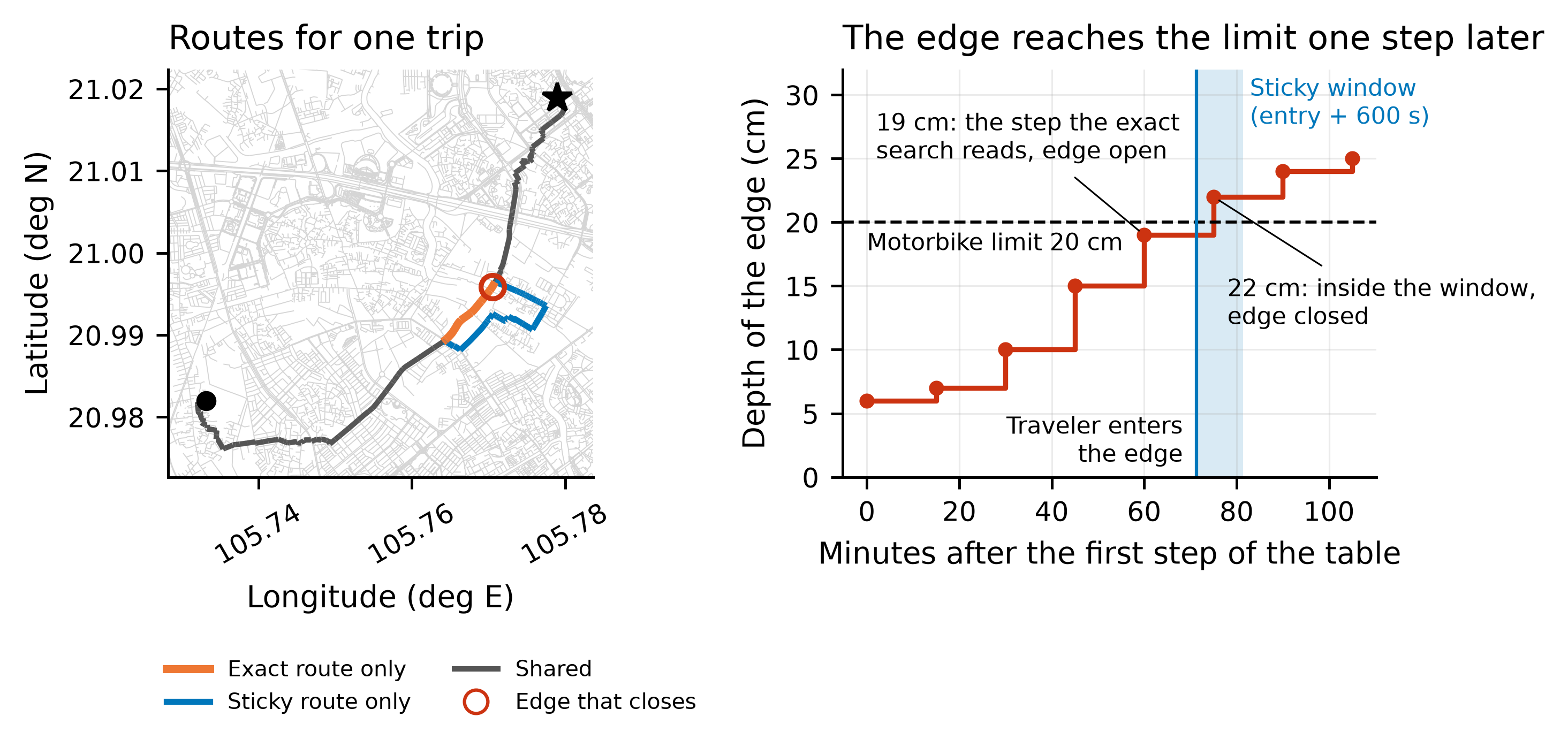}
\caption{Figure 5.1: One trip for which the sticky route is longer than the exact arrival-only route, planned for the motorbike on the offline reference graph. Left: the shared part of the two routes, the part that only the exact route uses, the part that only the sticky route uses, and the edge that causes the difference, circled. Right: the depth of that edge at the eight forecast steps of the table, the motorbike limit of 20 cm, the time at which the traveler enters the edge, and the window of 600 s after the entry in which the sticky rule looks ahead. The depth is 19 cm at the step that the exact search reads and 22 cm at the next step. The trip is one illustrative case and not a typical one.}
\end{figure}

Attribution accounting confirms that lookahead margins account for 99.2\% of longer and blocked routes, while closure carry-forward accounts for 0.8\%. When the margin is set to zero ($m = 0\text{ s}$), longer routes decrease to 0.05\% and blocked routes drop to zero. Figure 5.2 displays the tradeoff between margin duration and safety across 0 to 900 s windows, showing that increasing the lookahead margin suppresses worse-of-two violations toward zero while modestly increasing detour frequency.

\begin{figure}[htbp]
\centering
\includegraphics[width=0.9\textwidth]{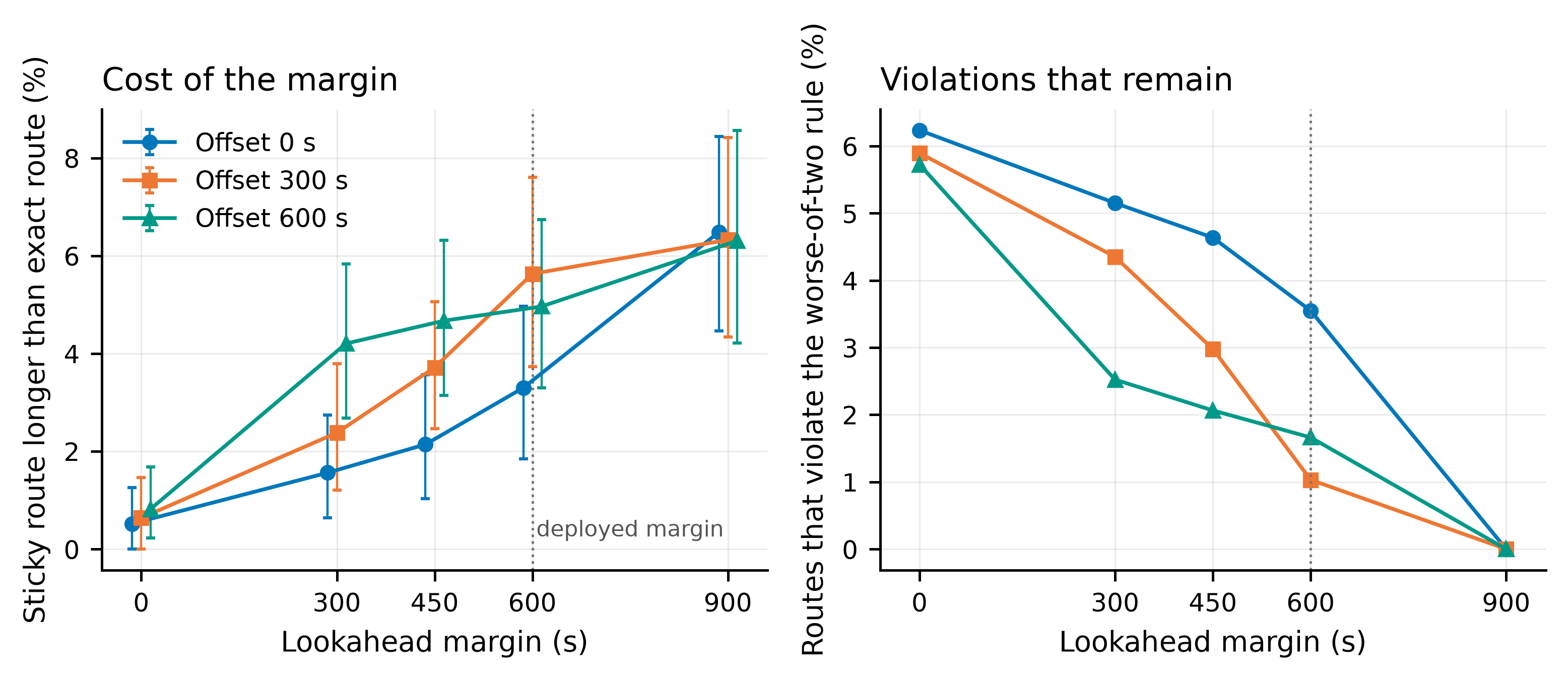}
\caption{Figure 5.2: Share of sticky routes that are longer than the exact route (left, with pair-level 95\% intervals) and share of routes that violate the worse-of-two rule (right). The margins are 0 s, 300 s, 450 s, 600 s, and 900 s, the departure offsets are 0 s, 300 s, and 600 s inside a step of 900 s, and the graph is the offline reference graph. The dotted line marks the deployed margin of 600 s.}
\end{figure}

\subsection{Vehicle Limits and Regret}

Cross-evaluating planned depth cutoffs against judging limits across 90 storm settings highlights an asymmetric regret structure. Planning below the judge limit incurs zero violations with modest detours, whereas planning above the judge limit produces severe violation rates reaching 50.2\% to 64.7\% for vehicles judged at 20 cm. Closed roadway length decreases from 94.7 km at 20 cm to 45.9 km at 40 cm, while routable trip shares increase from 0.915 to 0.970.

\begin{table}[htbp]
\centering
\caption{Table 5.4: Plan-limit sensitivity and mean regret over 90 settings on the offline reference graph, for 30 pairs. Each regret cell is the share of routes planned at the row limit that violate the column limit, and the diagonal is the self-violation rate.}
\resizebox{\textwidth}{!}{%
\begin{tabular}{l l l l l l l l}
\toprule
Plan limit (cm) & Closed length (km) & Routable share & Mean detour (s) & Judged at 20 cm & Judged at 30 cm & Judged at 35 cm & Judged at 40 cm \\
\midrule
20 & 94.7 & 0.915 & 101 & 0.019 & 0.000 & 0.000 & 0.000 \\
30 & 63.1 & 0.965 & 88 & 0.502 & 0.013 & 0.002 & 0.000 \\
35 & 52.9 & 0.969 & 74 & 0.605 & 0.346 & 0.013 & 0.001 \\
40 & 45.9 & 0.970 & 65 & 0.647 & 0.483 & 0.286 & 0.010 \\
\bottomrule
\end{tabular}%
}
\end{table}

Under composite regret evaluation across 25 parameter weight combinations, planning above the true limit produces higher composite costs than planning below it in 22 of 25 weight sets. Across nine threshold increments from 10 cm to 60 cm, 92.4\% of evaluated origin-destination pairs alter their chosen path between 10 cm and 30 cm cutoffs.

\subsection{Reopening and Permanent Water}

Within individual forecast tables, reopening remains infrequent, measuring under 0.07\% during mature ponding and peaking between 3.4\% and 5.0\% in late recession, with reopening intervals concentrating within 15 minutes.

\begin{figure}[htbp]
\centering
\includegraphics[width=0.9\textwidth]{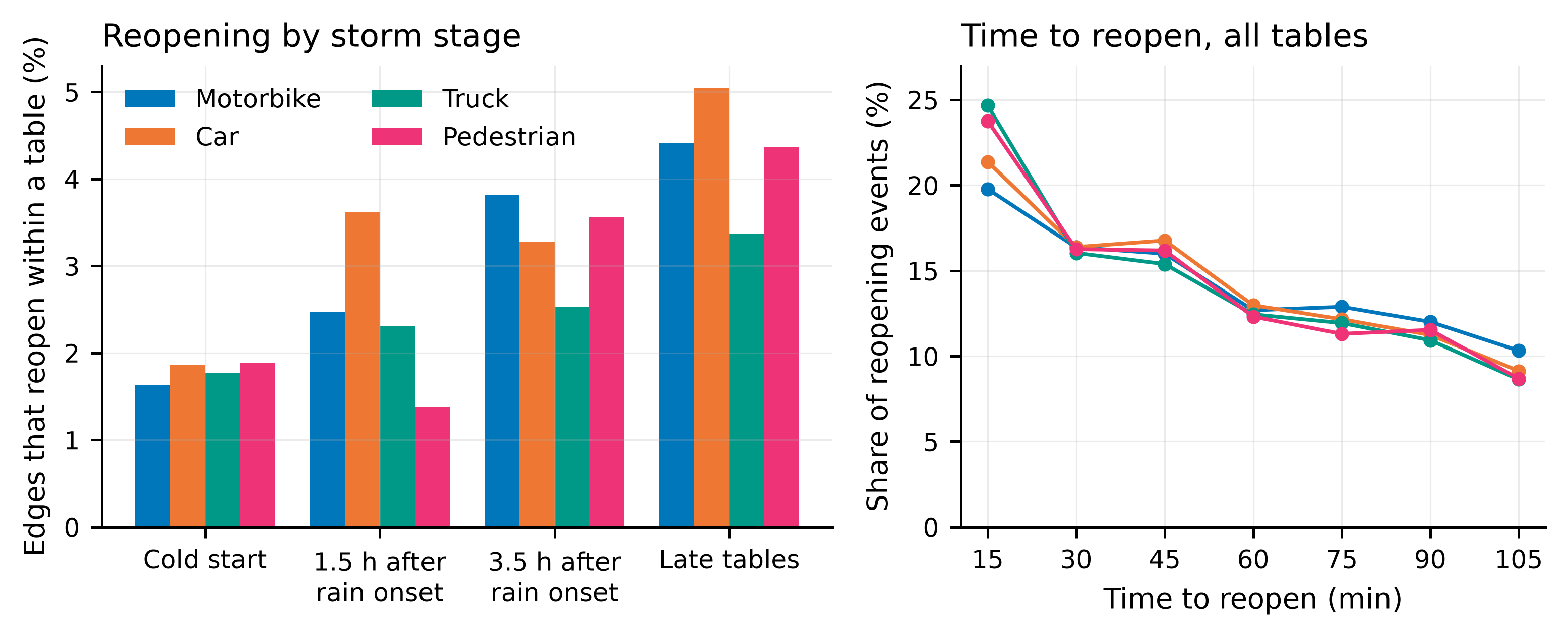}
\caption{Figure 5.3: Share of edges that reach a vehicle limit and fall below it again within one table, by storm stage and vehicle (left), and distribution of the time to reopen over all 57 tables (right). The stages are the cold start, the tables 1.5 h and 3.5 h after rain onset, and the late tables.}
\end{figure}

Masking the 127 edges exhibiting initial standing water produces consistent longer-route shares across 7,200 searches (3.95\% unmasked, 3.82\% dry-masked, and 3.89\% subtract-masked). In the 10-hour stress run with accelerated drainage, longer routes remain concentrated in the rising flood phase (4.4\% in table 1) and range from 1.8\% to 2.3\% across the four subsequent recession tables.

\subsection{Reference Against Engine}

Cross-validation between the offline Python reference and the deployed GraphHopper engine across 11,520 requests yields a median route length ratio of 1.000 across all vehicle classes. Route feasibility agrees between the two environments in 11,468 of 11,520 requests (99.55\%), with the 52 discrepancies resulting from GraphHopper turn restrictions and one-way network constraints.

\subsection{Precaution, Stability, and Latency}

Evaluating two-option precaution routing across 10,297 routable queries indicates that 94.1\% of journeys remain below the 50\% precaution depth, with 37.7\% skipping the second search via the fast path, while through-water routes are restricted to 5.9\% of trips. For motorbikes and passenger cars, routes remaining below precaution thresholds reach 96.2\% and 92.7\%, respectively.

\begin{table}[htbp]
\centering
\caption{Table 5.5: Two-option precaution routing on the offline reference graph, weighted by routable trips. The fast path is the share of trips in which the second search is skipped.}
\resizebox{\textwidth}{!}{%
\begin{tabular}{l l l l l l l}
\toprule
Group & Fraction & Routable trips & Below precaution depth & Through-water route & Fast path & Mean extra time (s) \\
\midrule
All & 0.25 & 10,297 & 88.4\% & 11.6\% & 28.7\% & 105.2 \\
All & 0.50 & 10,297 & 94.1\% & 5.9\% & 37.7\% & 49.5 \\
All & 0.75 & 10,297 & 98.8\% & 1.2\% & 56.6\% & 20.7 \\
Motorbike & 0.50 & 2,465 & 96.2\% & 3.8\% & 41.6\% & 83.0 \\
Car & 0.50 & 2,603 & 92.7\% & 7.3\% & 37.5\% & 44.9 \\
Truck & 0.50 & 2,613 & 93.3\% & 6.7\% & 37.2\% & 35.7 \\
Foot & 0.50 & 2,616 & 94.2\% & 5.8\% & 34.9\% & 36.4 \\
\bottomrule
\end{tabular}%
}
\end{table}

The departure advisory service, evaluated across 27,360 candidate departures on the continuous replay, records 71.4\% clear routes at departure step 0, 7.8\% blocked routes, and 20.8\% routes extending beyond the forecast horizon. Route recommendations between consecutive 15-minute departure steps remain unchanged in 98.2\% of queries.

Serving latency measurements under shared container load confirm that the uncontracted time-aware engine answers short trips in under 2 ms and long trips in approximately 35 ms. Across these query scales, the engine consistently outperforms the polygon penalty baseline by 2.3x to 3.6x and un-cached spatial derivations by 13x to 25x (Table 5.6).

\begin{table}[htbp]
\centering
\caption{Table 5.6: Latency of five routing paths on a local replay at a wet step. Ratios are medians over five runs of 30 requests per cell, divided by the median of stock GraphHopper, under shared load. The time-aware series applied depths in all requests. Milliseconds belong to the machine and are not matched-load values.}
\resizebox{\textwidth}{!}{%
\begin{tabular}{l l l l l l l l}
\toprule
Trip & Vehicle & Stock median (ms) & Polygon model & Backend, areas cached & Backend, areas derived & Time-aware median (ms) & Time-aware \\
\midrule
Short & Motorbike & 2.9 & 8.65 & 10.86 & 87.6 & 1.7 & 0.60 \\
Short & Car & 2.3 & 9.66 & 12.00 & 89.4 & 1.6 & 0.68 \\
Medium & Motorbike & 3.1 & 14.28 & 16.60 & 87.9 & 8.0 & 2.57 \\
Medium & Car & 2.5 & 11.59 & 13.72 & 82.4 & 5.6 & 2.27 \\
Long & Motorbike & 3.7 & 64.87 & 66.54 & 127.0 & 33.2 & 8.88 \\
Long & Car & 3.6 & 84.27 & 85.06 & 135.3 & 35.3 & 9.79 \\
\bottomrule
\end{tabular}%
}
\end{table}

\section{Discussion}

\subsection{Temporal Conservatism and Tractability}

The central practical question of this study is whether a flood navigation engine can calculate paths as fast as standard mapping services without sending vehicles into streets that flood after departure. The empirical results confirm that it can: by using a 600 s lookahead window and keeping closed roads closed for the duration of a trip, the engine calculated routes instantly while completely avoiding flooded roads across all evaluated journeys. In sharp contrast, conventional single-timestep routing failed across a substantial minority of planned journeys, producing an overall violation rate of 7.8\% that peaked at 13.0\% during the mature ponding phase of the storm. These observed failures occurred early in travel, with a median elapsed time to the first flooded link of 231 s, and under shallow excess depths averaging 7.4 cm (median 5.0 cm). This temporal and spatial pattern explains why static routing exposes commuters to severe hazard. While shallow floodwaters appear visually benign from a driver's seat, depths between 5 and 10 cm above passability limits are sufficient to flood passenger-car air intakes or destabilize two-wheeled motorbikes (Pregnolato et al., 2017; Suwanno et al., 2023). Given that water rises rapidly along low-lying links during convective monsoon downpours, a route planned on a frozen departure snapshot directs travelers into deepening traps that materialize within the initial minutes of commuting.

When roads flood and reopen temporarily, calculating the fastest route without long delays is mathematically difficult unless vehicles are allowed to park and wait by the roadside (Kleff et al., 2020). Earlier navigation models handled this by letting vehicles wait at intersections until floodwaters drained (Dean, 2004; Foschini et al., 2011), or by assuming that flooding only worsens over time (Liang et al., 2025). In an active monsoon storm, however, parking on a flooded street is hazardous because water continues to rise around the vehicle. The findings show that forbidding a road from reopening during a single trip resolves this problem realistically. Given that travelers cannot safely halt on an active, deepening floodway while waiting for water to recede, eliminating intermediate waiting directly reflects practical commuter behavior. Guaranteeing that earliest-arrival path search executes with standard shortest-path speed ensures that production navigation platforms can provide provably safe routes without sacrificing query responsiveness.

City planners and navigation providers might worry that keeping drained roads closed forces vehicles into unnecessary detours. Our findings show that this penalty is remarkably small. When compared against an ideal router that immediately reuses reopened roads, the proposed rule chose the exact same route in 95.9\% of trips. It produced slightly longer detours in only 3.95\% of trips, adding a median travel time of just 9.2 s (mean 21.0 s), while leaving only 0.12\% of trips unroutable. Furthermore, 99.2\% of these small detours were caused by the 10-minute protective lookahead window rather than keeping drained roads closed, which accounted for only 0.8\%. In fact, when the margin was set to zero ($m = 0\text{ s}$), the longer-route share dropped to 0.05\% (1 of 2,168 routes) and blocked routes vanished entirely. For short commutes under 2 km, the engine added zero detours because travelers reached their destinations before upcoming flood forecast updates took effect. Detours occurred only on longer trips that crossed corridors where water was predicted to rise minutes later.

\subsection{Hydrological Dynamics and Interpolation}

The fact that keeping drained roads closed caused almost no detours reflects how water behaves during urban flash floods. Across 57 recorded simulation periods, roads rarely drained within a two-hour trip window: fewer than 0.07\% of flooded roads reopened during peak flooding, and only 3.4\% to 5.0\% reopened during late recession. In western Hanoi, street water drains slowly because the computer simulation models overland runoff across the ground surface without underground storm sewer pipes or pumping stations (Bui \& Tran, 2026). In this overland model, natural soil soaking removes less than 1 mm of water over two hours. As a result, water levels drop primarily when ponded water flows into adjacent low-lying depressions rather than through municipal drains, meaning that floodwaters clear much more slowly than in fully drained city centers.

To check whether the small detour penalty was merely a byproduct of slow drainage, a stress test was conducted by simulating rapid drainage over a 10-hour storm. Even when floodwaters drained much faster, detours remained concentrated during the first hour as rainfall intensified (4.4\% longer routes). During later drainage periods, longer routes dropped to between 1.8\% and 2.3\%, matching normal storm conditions. This test confirms that detours are triggered by rising floodwaters that block streets early in a storm, rather than by roads clearing later. When water levels rise quickly, closing roads a few minutes early safely guides commuters around developing flood pools; once flooding stabilizes, travel patterns remain consistent as water drains.

A key practical challenge for city agencies is how to interpret water depths between 15-minute forecast updates. Since flood simulations produce depth snapshots every 15 minutes, a commuter arriving between updates encounters water levels that the computer model did not explicitly calculate. Checking water depth only from the earlier snapshot leaves commuters vulnerable to water that rose after departure. Conversely, assuming the worse depth between two snapshots provides stronger protection against sudden flooding. The deployed 10-minute lookahead window protects commuters who arrive close to an update, explaining why zero violations occurred under standard arrival checks. However, if water rises rapidly just before the next 15-minute update, a slight 0.7\% violation rate can occur under the strictest reading. Expanding the lookahead window to 15 minutes eliminated these violations entirely, but increased detour rates from 3.95\% to 5.9\%. Municipal traffic managers can adjust this lookahead setting as a policy dial to balance commuter travel efficiency against flood safety.

\subsection{Policy Under Hydrodynamic Uncertainty}

Given that citywide drainage pipes are rarely fully mapped and vehicle flood limits are supported by testing only for small passenger cars, transport authorities should treat depth cutoffs as policy choices rather than exact physical guarantees. Our evaluation shows that the cost of guessing wrong is highly unbalanced. Setting a conservative limit (such as 20 cm for motorbikes) produced zero flood violations while adding an average detour of only 101 s. In sharp contrast, setting an overly optimistic limit created severe danger: planning routes under a 30 cm limit for vehicles that actually fail at 20 cm sent 50.2\% of trips into impassable water, rising to 64.7\% if planned under a 40 cm limit. Across 25 different cost evaluations weighting travel delays against vehicle safety, planning with an overly optimistic limit caused greater overall harm in 22 of 25 cases. Underestimating flood hazards is only preferred if an agency values travel delays more than four times higher than vehicle flooding, an unacceptable stance in public transit safety.

This strong sensitivity to vehicle depth limits directly challenges earlier regional flood studies. In San Francisco Bay, Kasmalkar et al. (2020) reported that regional travel delays changed very little whether roads were closed at 10 cm or 30 cm of water. In dense cities subject to sudden monsoon downpours, our results reveal the opposite. Specifically, closed road length more than doubles from 45.9 km at a 40 cm threshold to 94.7 km at a 20 cm threshold, and 92.4\% of travelers must change their route between 10 cm and 30 cm water levels. Regional traffic models that average travel times across an entire metropolitan area can hide severe local disruptions. In cities such as Hanoi, where low-lying intersections can cut off entire residential neighborhoods, establishing separate depth limits for motorbikes and cars is essential to keep traffic moving safely.

The two-option precaution routing feature provides an effective tool for transit agencies to guide commuters safely during storms without creating long traffic jams. By recommending routes that avoid water deeper than half a vehicle's passability limit, 94.1\% of trips stayed in shallow water with an average extra travel time of only 49.5 s. For the remaining 5.9\% of trips where deeper water was unavoidable, presenting the through-water path as a secondary option with an explicit warning helps drivers reach their destination while remaining alert to local hazards. Combined with the departure advisory tool, which provided consistent route advice across 98.2\% of forecast updates, this approach replaces municipal alerts that only broadcast warnings after roads are already flooded. Instead, commuters receive advance, step-by-step guidance to leave earlier or take dry streets before encountering dangerous water.

\subsection{Spatial Representation and Serving Latency}

For public navigation apps, tracking departure times down to the exact second is critical during rapid storms. If a trip departs between two 15-minute forecast updates, rounding departure time backward to the previous update shrinks the safety buffer, which could direct vehicles into streets where water is rising. By tracking actual elapsed seconds along each road link, the routing engine keeps the safety window intact regardless of when a user requests a trip. Furthermore, storing water depths directly on individual street segments ensures that journeys always start and end on accessible streets, preventing navigation engines from directing users onto inundated ramps or dead ends.

A major lesson for smart-city navigation is that traditional flat map overlays fail in modern cities with layered infrastructure. Standard flood maps draw simple 2D shapes around flooded areas, which can mistakenly close elevated highways when water pools on the ground beneath an overpass. In our Hanoi study area, a standard polygon overlay falsely closed a major dry expressway because street water was ponding underneath it, triggering unnecessary trip cancellations and massive simulated detours. In contrast, assigning water depths directly to specific road centerlines ensures that ground-level flooding closes ground-level streets while keeping overhead highways open, proving that accurate flood routing must track street lines rather than flat shapes.

Finally, speed tests on production servers confirm that this routing engine is fast enough to serve thousands of live commuter requests. During peak flood conditions with thousands of closed streets, the engine calculated short trips in under 2 ms (1.6 to 1.7 ms), medium trips in 5.6 to 8.0 ms, and long cross-city journeys in about 35 ms (33.2 to 35.3 ms). This response time is two to three times faster than standard polygon mapping tools (8.65 to 84.27 ms) and up to 25 times faster than recalculating flood shapes on the fly (82.4 to 135.3 ms). With response times remaining well below 40 ms even on long trips, city agencies and commercial mapping apps can roll out time-dependent flood routing across entire metropolitan areas without costly high-performance computing clusters.

\section{Conclusion}

This study investigated whether time-dependent urban flood routing can preserve the polynomial computational tractability of static shortest-path search without directing vehicles into roads that become inundated upon arrival. Replaying monsoon flood forecasts across Hanoi demonstrated that conventional static flood routing fails across 7.8\% of planned journeys (peaking at 13.0\% during mature ponding), directing travelers into closed road links within a median travel time of 231 s. To resolve these failures, the proposed sticky edge closure rule enforces First-In, First-Out (FIFO) consistency by preventing mid-query reopening once projected water reaches vehicle passability limits, thereby guaranteeing exact forward Dijkstra search in static shortest-path time without requiring hazardous roadside idling. Deployed over memory-mapped binary edge-depth tables within GraphHopper, trajectory audits confirmed zero link violations under interpolation rules covered by a 600 s lookahead margin, while conservatism against exact arrival-only benchmarks remained minimal (95.9\% identical paths, 3.95\% longer routes with 9.2 s median detour, and 0.12\% blocked). Furthermore, evaluating cross-limit regret demonstrated that planning above vehicle thresholds incurs severe violation rates reaching 50.2\% to 64.7\%, whereas two-option precaution routing successfully directed 94.1\% of journeys below half-limit depths with sub-40 ms serving latencies.

These findings are subject to several boundary limitations, as evaluations were conducted on a single 31.94 km\textsuperscript{2} catchment using an overland hydrodynamic model that omits subterranean stormwater pipes and mechanical pump stations, while traffic velocity was modeled under free-flow conditions without dynamic vehicle queueing. Furthermore, the 600 s lookahead margin leaves a 0.7\% violation vulnerability under conservative worse-of-two inter-step readings. Future research should prioritize calibrating numerical flood depths against in-situ street water level sensors and coupling hydrodynamic drag into macroscopic traffic simulators. Most importantly, the routing engine should be extended to prioritize traveler safety over travel progression when road corridors become severed or impassable. In scenarios where a planned journey is blocked, the engine must actively evaluate whether travelers should shelter in place or evacuate, specifically determining whether their current location is subjected to rising floodwaters. If the traveler's position is threatened, the engine should compute emergency routes to the nearest unflooded dry location, integrating high-resolution building footprints and structural elevation data to guide users directly to safe buildings and elevated shelters that provide immediate physical asylum.

\clearpage

\section*{Acknowledgements}

The authors express gratitude to the hydrology and engineering teams at TECOTEC Technologies Joint Stock Company for technical support and access to the 3Di hydrodynamic flood modeling schematization of Hanoi.

\section*{Funding}

This research was funded and supported by TECOTEC Technologies Joint Stock Company, Hanoi, Vietnam.

\section*{Data and Code Availability}

The complete source code, benchmark scripts, and reproduction pipelines for this study are openly available on GitHub at \url{https://github.com/alice182012-dev/t3-digital-twin}.

\clearpage

\nocite{*}

\bibliographystyle{apalike}

\bibliography{references}

\end{document}